\documentclass[final,5p,times,twocolumn,nofootinbib,nopreprintline]{elsarticle}
\usepackage{amsmath,amssymb,amsfonts}
\usepackage{graphicx}
\usepackage{hyperref}
\usepackage{slashed}
\usepackage{booktabs,tabulary}
\usepackage{mciteplus}
\usepackage{color}

\usepackage{bibentry}
\newcommand{\ignore}[1]{}
\newcommand{\nobibentry}[1]{{\let\nocite\ignore\bibentry{#1}}}

\makeatletter
\def\bibinfo@X@title#1,{\ignorespaces}
\makeatother

\begin{document}
%%%%%%%%%%%%%%%%%%%%%%%%%%%%%%%%%%%%%%%%%%%%%%%%%%%%%%%%%%%%%%%%%%%%%%%%%%%%%

\begin{frontmatter}

\title{Dispersive analysis of the $\gamma\gamma^{*} \to \pi \pi$ process}
\author[Mainz]{Igor Danilkin}
\author[Mainz]{Marc Vanderhaeghen}
\address[Mainz]{Institut f\"ur Kernphysik \& PRISMA  Cluster of Excellence, Johannes Gutenberg Universit\"at,  D-55099 Mainz, Germany}

\begin{abstract}
We present a theoretical study of the $\gamma\gamma^{*} \to \pi^+\pi^-, \pi^0\pi^0$ processes from the threshold through the $f_2(1270)$ region in the $\pi\pi$ invariant mass. We adopt the Omn\`es representation in order to account for rescattering effects in both s- and d-partial waves. For the description of the $f_0(980)$ resonance, we implement a coupled-channel unitarity. The constructed amplitudes serve as an essential framework to interpret the current experimental two-photon fusion program at BESIII. They also provide an important input for the dispersive analyses of the hadronic light-by-light scattering contribution to the muon's anomalous magnetic moment.  
\end{abstract}
%\begin{keyword}
  %DRs\sep .... 
%
%\PACS 11.55.Fv\sep ....
%\end{keyword}
\end{frontmatter}

\section{Introduction}\label{intro}

The two-photon fusion reaction is a prime example where using S-matrix constraints, such as analyticity and unitarity one can make predictions, which serve as direct input into the Standard Model calculation of the hadronic light-by-light (HLbL) scattering contribution to the muon's anomalous magnetic moment $a_\mu$. The HLbL contribution is currently the largest source of uncertainty in this precision quantity, which at present shows a 3 - 4~$\sigma$ deviation between theory and experiment~\cite{Jegerlehner:2017gek,Keshavarzi:2018mgv}. Ongoing experimental programs at FERMILAB~\cite{Grange:2015fou} and J-PARC~\cite{Mibe:2010zz} aim to reach a fourfold increase in precision in the direct measurement of $a_\mu$. This prospect calls to reduce the theory uncertainty accordingly, which in turn critically entails to reduce the error on the HLbL contribution by a concerted theoretical and experimental effort. Experimentally, two-photon fusion reactions are studied at $e^+ e^-$ colliders. When both leptons in the process $e^+e^- \to e^+e^- X$ are detected in the final state, this reaction allows to access the two-photon fusion process $\gamma^*\gamma^* \to X$, where both photons have a spacelike virtuality. The dominant HLbL contributions to $a_\mu$ are coming from the production of the lightest pseudoscalar mesons $X=\pi^0,\eta,\eta'$. The next important contribution comes from pion pairs, which we consider in this paper. The first measurement of the $\gamma \gamma^\ast \to \pi^0 \pi^0$ process has been reported recently by the Belle Collaboration \cite{Masuda:2015yoh} for $Q^2$ in the region from 3.5 - 30 GeV$^2$. At small momentum transfers, the BESIII Collaboration is currently analyzing both $\pi^+ \pi^-$ and $\pi^0 \pi^0$ production in the $0.2~{\rm {GeV}}^2 \lesssim Q^2 \lesssim 2~{\rm {GeV}}^2$ range~\cite{Redmer:2017fhg}, corresponding with the most relevant kinematical region for quantifying the HLbL contribution to $a_\mu$.

Very close to threshold, the $\gamma\gamma \to \pi\pi$ process has been studied in $\chi$PT up to two-loop accuracy~\cite{Gasser:2005ud,Gasser:2006qa} as a tool to access pion polarizabilities. Such approaches fail however to describe the resonance region, which require resummation techniques to comply with exact unitarity~\cite{Oller:1997yg,Dai:2014zta,GarciaMartin:2010cw, Hoferichter:2011wk, Danilkin:2012ua}. Among those, the most established ones respect analyticity properties of the S-matrix \cite{GarciaMartin:2010cw, Hoferichter:2011wk, Danilkin:2012ua, Oller:2007sh, *Oller:2008kf}. The energy range of applicability of such dispersive techniques is typically limited by the inelastic contributions and inclusion of higher partial waves. Extending such dispersive techniques to the partial wave helicity amplitudes of the single virtual $\gamma \gamma^\ast \to \pi \pi$ process is not straightforward, as in addition to the well-known low-energy constraints, partial-wave amplitudes exhibit kinematic constraints. Therefore, so far, the dispersive analyses of $\gamma \gamma^\ast \to \pi \pi$ have been limited to the s-wave and single-channel description \cite{Moussallam:2013una, Colangelo:2017fiz,*Colangelo:2017qdm,*Colangelo:2014dfa} which only covers the $f_0(500)$ resonance region. The aim of this work is to extend the dispersive approach to the coupled-channel case by including $K\bar{K}$ intermediate states and to include for the first time the d-wave contribution, which allows for a full dispersive formalism through the prominent $f_2(1270)$ tensor meson region. This will allow for a validation of such approach by forthcoming BESIII data for the $\gamma \gamma^\ast \to \pi \pi$ reaction, which is a prerequisite for a data-driven approach in quantifying the uncertainty of the HLbL  contribution to $a_\mu$.

\section{Formalism}
\subsection{Kinematics and observables}

The two-photon fusion reaction $\gamma\gamma^*\to \pi\pi$ is a subprocess of the unpolarized single tagged process $e^+(k_1)e^{-}(k_2)\to e^{+}(k_1')e^{+}(k_2') \pi(p_1)\pi(p_2)$ which is given (in Lorenz gauge) as
\begin{eqnarray}\label{ee->eepipi_1}
i\,{\cal M}&=&\frac{i\,e^2}{q_1^2 q_2^2}\,[\bar{v}(k_1)\,\gamma_\mu\,v(k_1')]\,[\bar{u}(k_2')\,\gamma_\nu\,u(k_2)]\,H^{\mu\nu}\,,\\
H^{\mu\nu}&=&  i\int d^4 x\,e^{-i\,q_1\cdot x} \langle \pi(p_1)\pi(p_2)|T(j^{\mu}_{em}(x)
\,j^{\nu}_{em}(0))|0\rangle\,, \nonumber
\end{eqnarray}
where the lepton momentum $k^\prime_2$ is detected, whereas the second lepton momentum $k^\prime_1$ goes undetected. This corresponds with the kinematical situation where the photon with momentum $q_2$ has a finite virtuality $q_2^2=-Q_2^2\equiv Q^2$, while the first photon with momentum $q_1$ is quasi-real, i.e. $q_1^2=-Q_1^2 \simeq 0$. The hadron tensor $H^{\mu\nu}$ satisfies gauge invariance, i.e. $q_{1\mu}\,H^{\mu\nu}=q_{2\nu}\,H^{\mu\nu}=0$, and can be expanded in terms of a complete set of invariant amplitudes \cite{Moussallam:2013una, Tarrach:1975tu,*Drechsel:1997xv}

\begin{equation}
H^{\mu\nu}=\sum_{i=1}^{3}F_i\,L^{\mu\nu}_i\,,
\end{equation}
where
\begin{align}
\label{Eq:Lorez_structures}
&L_1^{\mu\nu}=q_1^{\nu}\,q_2^{\mu}-(q_1\cdot q_2)\,g^{\mu\nu},\\
&L_2^{\mu\nu}=(\Delta^2\,(q_1\cdot q_2)-2\,(q_1\cdot
\Delta)\,(q_2\cdot
\Delta))\,g^{\mu\nu}-\Delta^2\,q_1^\nu\,q_2^\mu
  \nonumber\\  
  & -2\,(q_1\cdot q_2)\,\Delta^{\mu}\,\Delta^{\nu}
    +2\,(q_2\cdot \Delta)\,q_1^{\nu}\,\Delta^{\mu}+2(q_1\cdot \Delta)\,q_2^{\mu}\,\Delta^{\nu}\,,\nonumber\\
%%%
&L_3^{\mu\nu}=-\left(q_1\cdot \Delta\right)\left(g^{\mu \nu}Q^2+q_2^{\mu}q_2^{\nu}\right)
+\Delta ^{\mu}\left(q_2^{\nu}\left(q_1\cdot q_2 \right) +q_1^\nu Q^2\right),
\nonumber
\end{align}
with $\Delta \equiv p_1-p_2$, and where the numbering is chosen such that in the real photon case only $L^{\mu\nu}_{1,2}$ contribute. The latter coincide with the tensor structures used in \cite{Danilkin:2012ua, Danilkin:2017lyn,*Deineka:2018nuh}. The invariant amplitudes $F_i$ depend on the Mandelstam variables $s=(q_1+q_2)^2$, $t=(p_1-q_1)^2$ and $u=(p_1-q_2)^2$ which satisfy the relation $s+t+u=2\,m_{\pi}^2-Q^2$. One can notice that the tensor $L_3^{\mu\nu}$ is odd under pion crossing $(\Delta \to -\Delta)$. Therefore the amplitude $F_3$ is odd under $t \leftrightarrow u$ interchange and has a zero when $t=u$. As it was pointed out in \cite{Moussallam:2013una, Colangelo:2015ama}, this kinematic zero can be absorbed by the redefinition $L^{\mu\nu}_3 \to (t-u)\,L^{\mu\nu}_3$. We emphasize that the basis (\ref{Eq:Lorez_structures}) is minimal and non-degenerate in any kinematic point. The invariant amplitudes are free from kinematic singularities or constraints except for the Born terms, which are known to have a double pole structure in the soft-photon limit, as a manifestation of Low's theorem \cite{Low:1958sn}.

By contracting the hadronic tensor $H^{\mu\nu}$ with polarization vectors, one defines helicity amplitudes which can be further partial wave projected as\footnote{Note, that we use an extra factor $N=1/\sqrt{2} $ for p.w. expansion of the $\gamma\gamma^{*} \to K\bar{K}$ amplitudes (in contrast to  $N=1$ for $\gamma\gamma^{*} \to \pi\pi$) in order to match our normalization for the hadronic p.w. amplitudes, which ensure the same unitarity relations for the identical and non-identical particles.}
\begin{eqnarray}\label{p.w.expansion}
&&\epsilon_\mu(q_1,\lambda_1)\,\epsilon_\nu(q_2,\lambda_2)\,H^{\mu\nu}\equiv e^{i\phi(\lambda_1-\lambda_2)}H_{\lambda_1 \lambda_2}\\&&=e^{i\phi(\lambda_1-\lambda_2)} N \sum_{J}(2J+1)\,h^{(J)}_{\lambda_1\lambda_2}(s,Q^2)\,d_{\Lambda,0}^{(J)}(\theta)\,,\nonumber
\end{eqnarray}
where $\Lambda=\lambda_1-\lambda_2$, $d_{\Lambda,0}^{(J)}(\theta)$ is a Wigner rotation function and $\theta$ is the c.m. scattering angle. The two-photon initial state implies that the $C$-parity quantum number of the final particles should always be positive. This excludes the isospin $I=1$ state in the case of two pions, and due to Bose symmetry, only even values of total angular momentum $J$ survive in the p.w. expansion. Therefore, two photon fusion reactions provide valuable information on the nature of the scalar $f_0(500)$, $f_0(980)$ and  tensor $f_2(1270)$ resonances. We will work in the isospin limit, defining helicity amplitudes $H_{I, \lambda_1\lambda_2}$ ($K_{I, \lambda_1\lambda_2}$) for 
$\gamma\gamma^*\to \pi \pi$ ($K \bar K$), 
which imply the following relations for $I=0$ and $I=2$
\begin{align}
&H_{0,\lambda_1\lambda_2}=-\frac{2 H_{\lambda_1\lambda_2}^c+H_{\lambda_1\lambda_2}^n}{\sqrt{3}}\,,~ K_{0,\lambda_1\lambda_2}=-\frac{K_{\lambda_1\lambda_2}^c+K_{\lambda_1\lambda_2}^n}{\sqrt{2}}\,, \nonumber \\
&H_{2,\lambda_1\lambda_2}=\sqrt{\frac{2}{3}} \left(H_{\lambda_1\lambda_2}^n-H_{\lambda_1\lambda_2}^c\right),
\end{align}
where $H_{\lambda_1\lambda_2}^c(K_{\lambda_1\lambda_2}^c)$ and $H_{\lambda_1\lambda_2}^n(K_{\lambda_1\lambda_2}^n)$ are the corresponding amplitudes for charged or neutral pion (kaon) pairs.

The helicity amplitudes are expressed in terms of the invariant amplitudes through the following form
\begin{align}
\label{HelAmpl_InvAmpl}
&H_{++}=\left(s+Q^2\right)\left(-\frac{F_1}{2}+\frac{2\,p^2}{s}Q^2  z^2 (F_2+(s+Q^2)\,F_3)\right)\,, \nonumber\\
&H_{+-}=\left(s+Q^2\right)\left(-2\,(1-z^2)\, p^2\,F_2\right)\,,\\
&H_{+0}=\left(s+Q^2\right)\sqrt{\frac{2\,Q^2}{s}}z\sqrt{1-z^2}\,p^2 (-2F_2-(s+Q^2)F_3)\,,\nonumber
\end{align}
where $z=\cos\theta$, and where the initial and final relative momenta in the c.m. frame are given by
\begin{equation}
q=\frac{s+Q^2}{2\sqrt{s}}\,,\quad p=\frac{1}{2}\sqrt{s-4\,m_\pi^2}=\frac{1}{2}\sqrt{s}\,\beta_{\pi\pi}\,.\nonumber
\end{equation}
We see from Eq.(\ref{HelAmpl_InvAmpl}) that when $s=-Q^2$ (and $t=u=m_\pi^2$) all helicity amplitudes are equal to zero except for the Born amplitudes which have an additional pole at this kinematical point. 

From the helicity amplitudes it is then straightforward to obtain the differential cross section as
\begin{align}\label{Eq:Cross_section}
&\frac{d \sigma_{TT}}{d \cos\theta}=\frac{\beta_{\pi\pi}}{64\,\pi\,(s+Q^2)}\left(|H_{++}|^2+|H_{+-}|^2\right)\,,\quad\\
&\frac{d \sigma_{TL}}{d \cos\theta}=\frac{\beta_{\pi\pi}}{32\,\pi\,(s+Q^2)}\, |H_{+0}|^2\,,\nonumber
\end{align}
where $\sigma_{TT}(\sigma_{TL})$ corresponds to the cross sections which involve two transverse photon polarizations (or when one of them is longitudinal with the polarization vector defined as $\epsilon_\nu(q_2,0) = 1/\sqrt{Q^2}\,\{-q, 0, 0, \sqrt{s}-q\}$).

\subsection{Dispersion relations}
In order to write down dispersion relations (DRs) for the $\gamma\gamma^* \to \pi\pi$ process, one has to identify all the kinematic constraints of the p.w. helicity amplitudes. While for the case of the on-shell photons, helicity amplitudes are not correlated at any kinematic point, this is no longer the case for $Q^2 \neq 0$. It can be most easily seen by expressing invariant amplitudes $F_{1,2,3}$ in terms of the p.w. helicity amplitudes. For the s-wave ($J_{max}=0$ in the p.w. expansion of Eq. (\ref{p.w.expansion})) one obtains as contributions
\begin{equation}
F_1=-\frac{2\, h_{++}^{(0)}}{\left(s+Q^2\right)}\,,\quad F_2=F_3=0\,,\nonumber
\end{equation}
while for $J_{max}=2$, one obtains the contributions
\begin{align}
F_2=\frac{-5 \sqrt{\frac{3}{2}}\, h_{+-}^{(2)}}{\left(s-4\, m_\pi^2\right) \left(s+Q^2\right)}\,, F_3=\frac{5\,\sqrt{6}\, \left(h_{+-}^{(2)}-\sqrt{2s/Q^2}\, h_{+0}^{(2)}\right)}{\left(s-4\,m_\pi^2\right) \left(s+Q^2\right)^2}\,,\nonumber
\end{align}
with a lengthy expression for $F_1$ that involves $h_{++}^{(0)}$, $h_{++}^{(2)}, h_{+-}^{(2)}$, $h_{+0}^{(2)}$ and angular dependencies. One notices that at finite $Q^2$ the s-wave contribution is not correlated with any other p.w. amplitude and one can write a DR by just accounting for the overall factor $(s+Q^2)$ which is required by the soft-photon theorem, i.e. 
\begin{equation}
h^{(0)}_{++}-h^{(0), Born}_{++} \simeq (s+Q^2)\,.
\end{equation}
The same holds for the helicity-2 p.w. amplitudes where one can identify the kinematic factors at low energies as
\begin{equation}\label{Constraints_hel2}
h^{(J)}_{+-}-h^{(J), Born}_{+-} \simeq (s+Q^2)\,p^J\,q^{J-2}\,.
\end{equation}
In contrast, the helicity amplitudes $h^{(2)}_{+-}(s)$ and $h^{(2)}_{+0}(s)$ are linearly dependent at $s=-Q^2$ even after accounting for the corresponding overall factors, as in Eq.(\ref{Constraints_hel2}). This problem was discussed in detail in \cite{Lutz:2011xc, Gasparyan:2010xz} and can be fixed by working with a kinematically unconstrained basis. The transformation matrix between different bases can be obtained by analyzing projected helicity amplitudes in terms of the quantities,
\begin{equation}
A_n^J(s)=\frac{1}{(p\,q)^J}\int_{-1}^{1}\frac{dz}{2}P_J(z)\,F_n(s,t)\,,
\end{equation}
which are free of any singularities due to the properties of the Legendre polynomials \cite{Lutz:2011xc}. It follows that the set of amplitudes $\{{\bar{h}}^{(J)}_{I,1},{\bar{h}}^{(J)}_{I,2},{\bar{h}}^{(J)}_{I,3}\}$, defined in terms of $\{\bar{h}^{(J)}_{I,+-},\bar{h}^{(J)}_{I,+0},\bar{h}^{(J)}_{I,++}\}$ as
\begin{eqnarray}\label{BarierFactors_Qneq0}
\left(\begin{array}{c}
{\bar{h}}^{(J)}_{I,1}\\
{\bar{h}}^{(J)}_{I,2}\\
{\bar{h}}^{(J)}_{I,3}
\end{array}
\right)&=&\frac{1}{(s+Q^2)\,p^J\,q^{J-2}}\,\,\textbf{M}
\left(
\begin{array}{c}
\bar{h}_{I,+-}^{(J)}\\
\bar{h}_{I,+0}^{(J)}\\
\bar{h}_{I,++}^{(J)} 
\end{array}
\right)\,,\\
\textbf{M}&=&\left(
\begin{array}{ccc}
 1 & 0 & 0 \\
 \frac{1}{\beta_J}\frac{1}{s+Q^2} & -\frac{1}{\sqrt{2}\,\gamma_J}\frac{1}{s+Q^2}\sqrt{\frac{s}{Q^2}} & 0 \\
 -\frac{\alpha_J}{\beta_J}\frac{Q^2}{s\,q^2} & \frac{\sqrt{2}\,\alpha_J}{\gamma_J}\frac{Q^2}{s\,q^2}\sqrt{\frac{s}{Q^2}} & \frac{1}{q^2} \\
\end{array}
\right)\,,\nonumber
\end{eqnarray}
are free from any constraints. In Eq.(\ref{BarierFactors_Qneq0}) $\bar {h}^{(J)}_{I}$ stand for the Born subtracted amplitudes and $\alpha_J,\,\beta_J,\,\gamma_J$ are numeric factors which for $J=2$ correspond to $\left\{\alpha _{2}, \beta _{2}, \gamma _{2}\right\}=\left\{2/15, 2/5\sqrt{2/3}, 1/5\sqrt{2/3}\right\}$. After identifying all the kinematic constraints, we are now in a position of constructing DRs, which unitarize our p.w. amplitudes. The photon fusion amplitudes $\gamma\gamma^* \to \pi\pi, K\bar{K}$ (or $\gamma\gamma^* \to \pi\pi$) are the off-diagonal elements of the $(\gamma\gamma^*), (\pi \pi), (K\bar{K})$ channels. Since the intermediate states with two photons are proportional to $e^4$, they are suppressed, and one can reduce the $(3\times 3)$ matrix DR down to the $(2\times1)$ form, which require the hadronic rescattering part as input. The unitarity relation for $s \geq 4\,m_\pi^2$ can be written as
\begin{align}\label{h_Unitarity}
&\left(\begin{array}{c}\text{Disc}\,h^{(J)}_{I,++}\\\text{Disc}\,k^{(J)}_{I,++}\end{array}\right)=t^{(J)*}_{I}\,\rho\,
\left(\begin{array}{c} h^{(J)}_{I,++}\\ k^{(J)}_{I,++}\end{array}\right)\,,\quad \\
&\rho=\frac{1}{16\,\pi}\left(
\begin{array}{cc}
\beta_{\pi\pi}(s)\,\theta(s-4m_\pi^2)& 0\\
0 & \beta_{KK}(s)\,\theta(s-4m_K^2)
\end{array}
\right)\,,\nonumber
\end{align}
where $\rho(s)$ is a two-body phase space factor and $t^{(J)}_{I}(s)$ is the coupled-channel $\{\pi\pi, K\bar{K}\}$ scattering amplitude, which is normalized as $\text{Im }(t^{(J)}_{I})^{-1}=-\rho$.
For the s-wave we write an unsubtracted DR for the function $(h^{(0)}_{I,++}-h^{(0), Born}_{I,++})\,(\Omega_I^{(0)})^{-1}/(s+Q^2)$, which contains both right and left hand cuts. This particular separation of the left-hand cuts into Born and non-Born parts was first used in \cite{GarciaMartin:2010cw} for the real photon case. The so-called Omn\`es function satisfies the following unitarity constraint above the two-pion threshold
\begin{equation}\label{Omnes_CC_Unitarity}
\text{Disc}\,\Omega^{(J)}_{I}=t^{(J)}_{I}\rho\,\Omega^{(J)*}_{I}\,.
\end{equation} 
For a proper description of the $f_0(980)$ resonance we employ for $I=0$ the coupled-channel equation
\begin{align}\label{Rescattring:I=0}
&\left(\begin{array}{c}
h^{(0)}_{0,++}\\
k^{(0)}_{0,++}\end{array}\right)=\left(\begin{array}{c}h^{(0),Born}_{0,++}\\ k^{(0), Born}_{0,++}\end{array}\right)+(s+Q^2)\,\Omega^{(0)}_{0}(s)\\
&\quad\quad \times \left[-
\int_{4m_\pi^2}^{\infty}\frac{ds'}{\pi}\,\frac{\text{Disc}\,(\Omega^{(0)}_{0}(s'))^{-1}}{(s'+Q^2)\,(s'-s)}\left(\begin{array}{c}h^{(0),Born}_{0,++}(s')\\ k^{(0),Born}_{0,++}(s')\end{array}\right)\right. \nonumber \\
&\quad\quad\quad\quad +
\left.\int_{-\infty}^{s_L}
\frac{ds'}{\pi}\,\frac{(\Omega^{(0)}_{0}(s'))^{-1}}{(s'+Q^2)\,(s'-s)}\left(\begin{array}{c}\text{Disc}\,\bar{h}^{(0)}_{0,++}(s')\\\text{Disc}\,\bar{k}^{(0)}_{0,++}(s')\end{array}\right)
\right]\,,\nonumber
\end{align}
with
\begin{equation}\label{Omnes_CC}
\Omega_0^{(0)}(s)=\left(
\begin{array}{cc}
\Omega(s)_{\pi\pi \to\pi\pi} & \Omega(s)_{\pi\pi \to K\bar{K}}\\
\Omega(s)_{K\bar{K} \to \pi\pi} & \Omega(s)_{K\bar{K} \to K\bar{K}}
\end{array}
\right)\,,
\end{equation}
while for $I=2$ we use a single-channel version of it. In (\ref{Rescattring:I=0}) both p.w. amplitudes and $s_L$ have a $Q^2$ dependence. For $J=2$ we write the set of single-channel dispersion integrals for $\left({h}^{(2)}_{I,i}-{h}^{(2), Born}_{I,i}\right)\,(\Omega^{(2)}_{I})^{-1}
$ which leads to
\begin{align}\label{Rescattring:general}
&{h}^{(2)}_{I,i}= {h}^{(2),Born}_{I,i}+ 
%\\ &&
\Omega^{(2)}_{I}(s) \bigg[
-\int_{4m_\pi^2}^{\infty}\frac{ds'}{\pi}\,\frac{\text{Disc}\,(\Omega^{(2)}_{I}(s'))^{-1}\,{h}^{(2),Born}_{I,i}(s')}{(s'-s)}
\nonumber \\
& \hspace{2cm}+
\int_{-\infty}^{s_L}
\frac{ds'}{\pi}\,\frac{(\Omega^{(2)}_{I}(s'))^{-1}\,\text{Disc}\,{\bar{h}}^{(2)}_{I,i}(s')}{(s'-s)}
\bigg]\,,
\end{align}
where $i=1,2,3$ and $s_L$ defines the position of the left-hand singularity nearest to the physical region due to non-Born intermediate $t$ - and $u$-channel left-hand cuts. 

\subsection{Right- and left-hand cuts}
To evaluate the DRs of Eqs.~(\ref{Rescattring:I=0}) and (\ref{Rescattring:general}), we need to specify the right- and left-hand cuts. 

The right-hand cut is fully specified through the Born amplitude and the hadronic Omn\`es function.
For the d-wave $I=0,2$ amplitudes we use the single-channel Omn\`es function given in terms of the corresponding phase shifts,
\begin{equation}\label{OmenesPhaseShift}
\Omega_I^{(2)}(s)=\exp\left(\frac{s}{\pi}\int_{4m_\pi^2}^{\infty} \frac{d s'}{s'}\frac{\delta_{I}^{(2)}(s')}{s'-s}\right).
\end{equation}
Its numerical evaluation requires a high-energy parametrization of the phase shifts. We use a recent Roy analysis \cite{GarciaMartin:2011cn} below 1.42 GeV, and let the phase smoothly approach $\pi$ ($0$) for $I=0$ ($I=2$) respectively. For the s-wave $I=0$ amplitude we use the coupled-channel Omn\`es function from a dispersive summation scheme \cite{Gasparyan:2010xz, Danilkin:2010xd} which implements constraints from analyticity and unitarity. The method is based on the $N/D$ ansatz \cite{Chew:1960iv}, where the set of coupled-channel integral equations for the $N$-function are solved numerically with the input from the left-hand cuts which we present in a model-independent form as an expansion in a suitably constructed conformal mapping variable. These coefficients in principle can be matched to $\chi$PT at low energy \cite{Danilkin:2011fz,*Danilkin:2012ap}. Here we use a more data-driven approach, and determine these coefficients directly from fitting to Roy analyses for $\pi\pi \to \pi\pi$ \cite{Ananthanarayan:2000ht,GarciaMartin:2011cn}, $\pi\pi \to K\bar{K}$ (for $I=0$) \cite{Buettiker:2003pp,Pelaez:2018qny} and existing experimental data for these channels. After solving the linear integral equation for $N(s)$, the $D$-function (the inverse of the Omn\`es function) was computed. Details will be given elsewhere \cite{Danilkin:2018}. The partial waves beyond $s$- and $d$-waves are approximated by the Born terms.

%For point-like pions (kaons), the dominant left-hand cuts can be described in terms of the scalar QED Lagrangian supplemented with light vector meson degrees of freedom \cite{GarciaMartin:2010cw,Danilkin:2012ua}. However, this is not enough for the case of virtual photons where the finite-size of the particles requires the introduction of corresponding form factors. 
We start with the most important contribution which comes from the pion (kaon) pole. The off-shellness of the photon can be taken into account through the pion (kaon) vector form factor, $f_{\pi,K}(Q^2)$, which is determined as a matrix element of the EM current between the two on-shell pions (kaons)
\begin{equation}
\langle \pi^+(p') | j_{\mu }(0) | \pi^+(p)\rangle =e \left(p+p'\right)_{\mu}f_{\pi }\left((p'-p)^2\right). \nonumber
%\\ \langle K^+(p') | j_{\mu }(0) | K^+(p)\rangle &=& \left(p+p'\right)_{\mu}f_{K}\left((p'-p)^2\right)\nonumber
\end{equation}
It was shown in \cite{Colangelo:2015ama} (see also \cite{Fearing:1996gs}) using the fixed-$s$ Mandelstam representation, that the pion pole contribution coincides exactly with the scalar QED Born contribution multiplied by the electromagnetic pion form factors. The invariant amplitudes due to these pole (Born) contributions are given by
\begin{align}\label{Fi:Born}
&F_1^{Born}=-\frac{e^2 \left(4\,m_i^2+Q^2\right)}{\left(t-m_i^2\right) \left(u-m_i^2\right)}\,f_{i}(Q^2)\,,\quad \\
&F_2^{Born}=-\frac{e^2}{\left(t-m_i^2\right) \left(u-m_i^2\right)}\,f_{i}(Q^2)\,,\quad 
F_3^{Born}=0, \nonumber
\end{align}
where $i={\pi,K}$. We note that the double-pole structure of the Born helicity amplitudes brings an additional kinematic constraint to the p.w. amplitudes. However, for spacelike photons, the point $s=-Q^2$ lies in the unphysical region and does not bring any additional complication. In contrast, for the case of timelike photons, this pole singularity may overlap with the unitary cut when $-Q^2=q^2>4m_\pi^2$ and an appropriate analytic continuation is necessary $q^2 \to q^2+i\epsilon$ \cite{Moussallam:2013una}. In this work, we parametrize the spacelike pion and kaon electromagnetic form factors by a simple monopole form 
\begin{equation}
f_{\pi,K}(Q^2)=\frac{1}{1+Q^2/\Lambda_{\pi,K}^2}\,,
\end{equation}
which provides a good description of the $Q^2 \lesssim 1$ GeV$^2$ data both for the pion form factor~\cite{Ackermann:1977rp,*Dally:1981ur,*Amendolia:1986wj} and kaon form factor \cite{Dally:1980dj,*Amendolia:1986ui}. The resulting values for the mass parameters are
$\Lambda_{\pi}=0.727(5)$ GeV and $\Lambda_K=0.872(47)$ GeV with $\chi^2/\text{d.o.f.}=1.22$ and $\chi^2/\text{d.o.f.}=0.69$, respectively. 

\begin{figure}[t]
\includegraphics[width =0.475\textwidth]{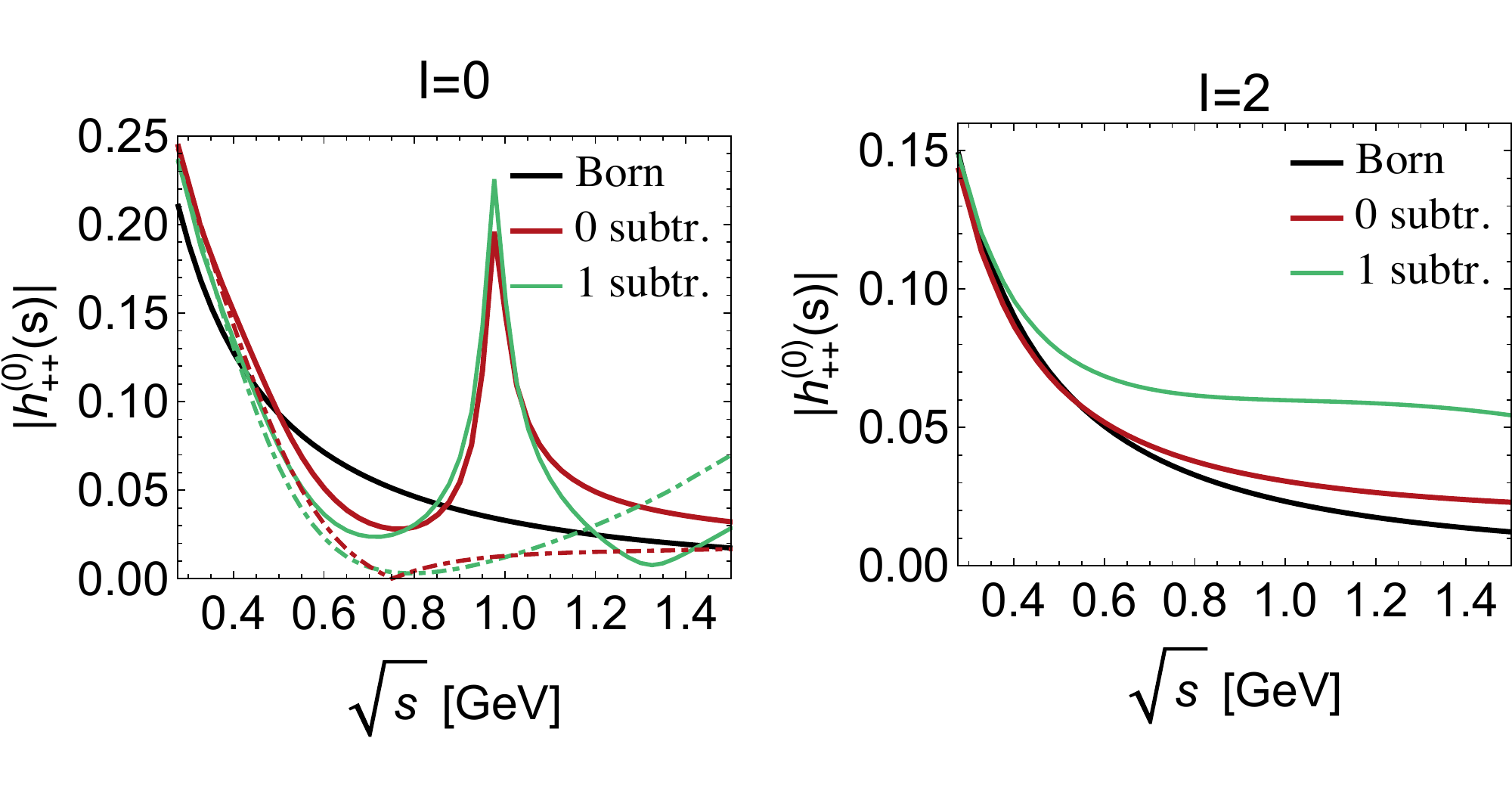}
\caption{The moduli of the s-wave helicity amplitudes for $I=0$ and $I=2$. Results of the dispersive analyses with 0 (1) subtractions are shown by the red (green) curves. In the case of $I=0$ the single- (coupled-) channel results are shown by the dashed-dotted (solid) curves. Black solid curves: Born result. }
\label{Fig:I=0}
\end{figure}

\begin{figure*}[t]
\centering
\includegraphics[width =0.37\textwidth]{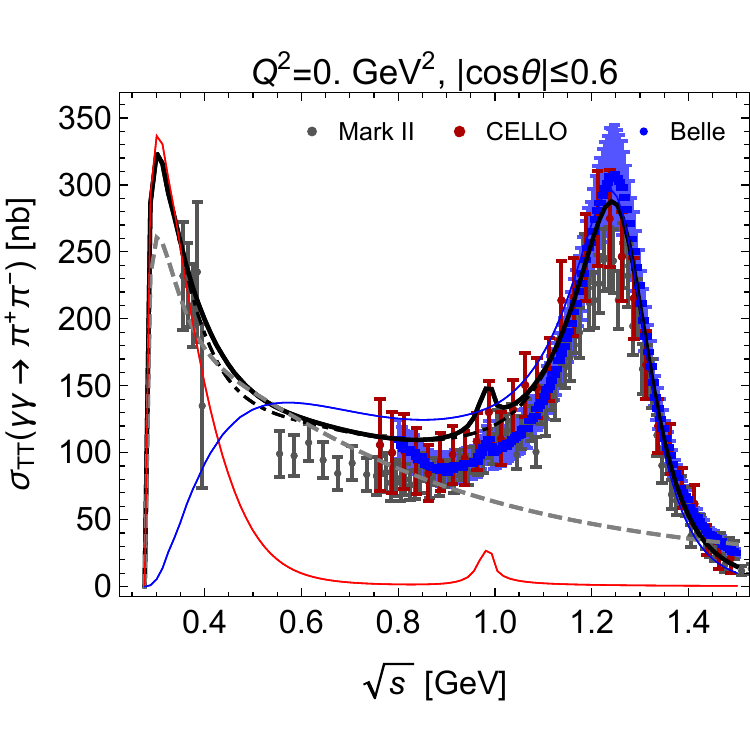}\quad \includegraphics[width =0.46\textwidth]{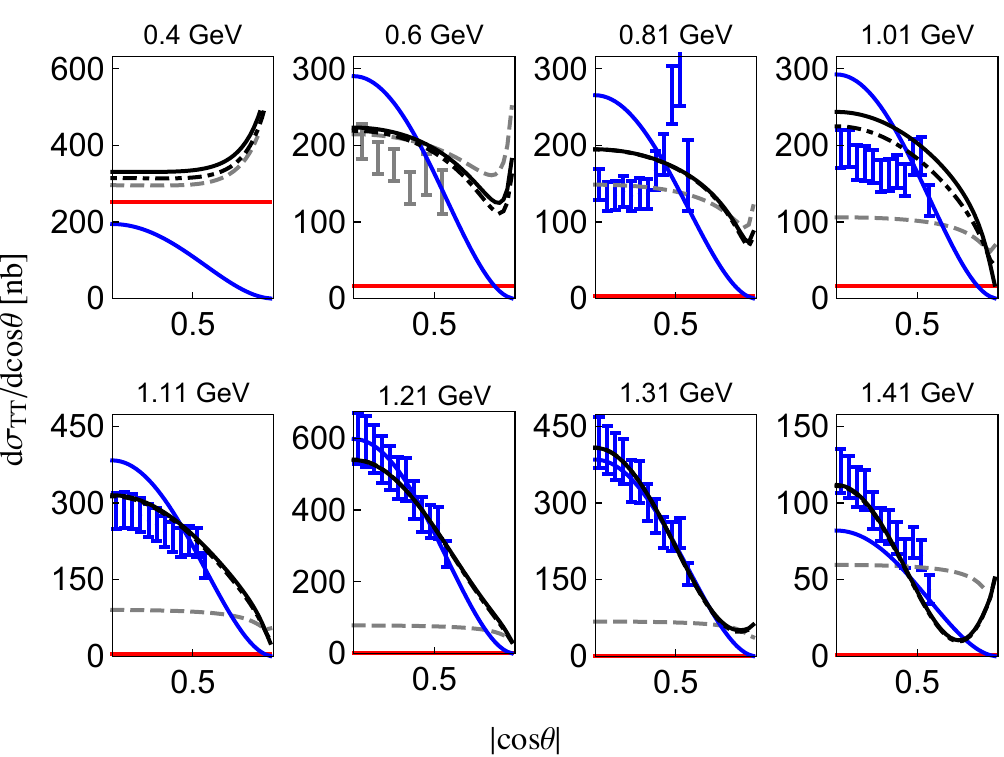}\\
\includegraphics[width =0.37\textwidth]{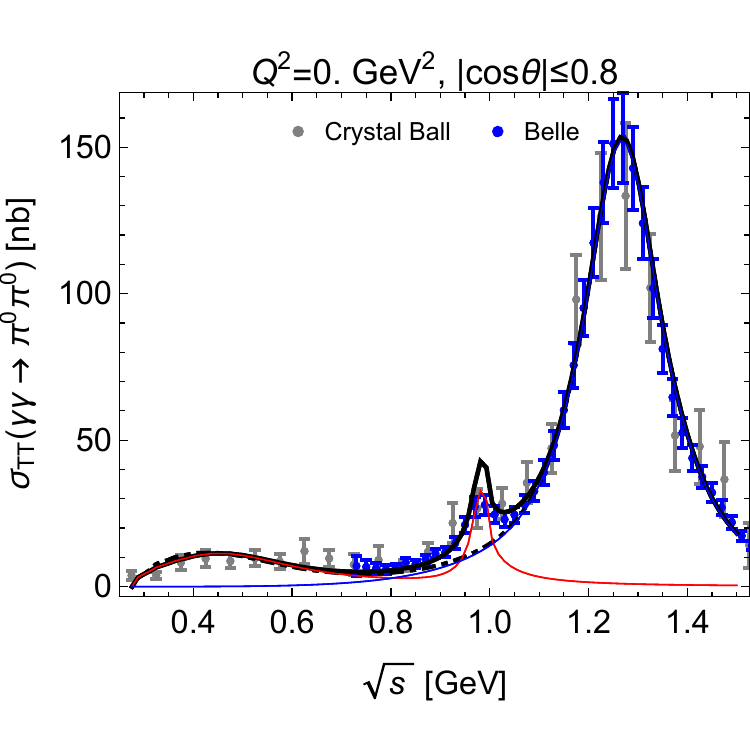}\quad \includegraphics[width =0.46\textwidth]{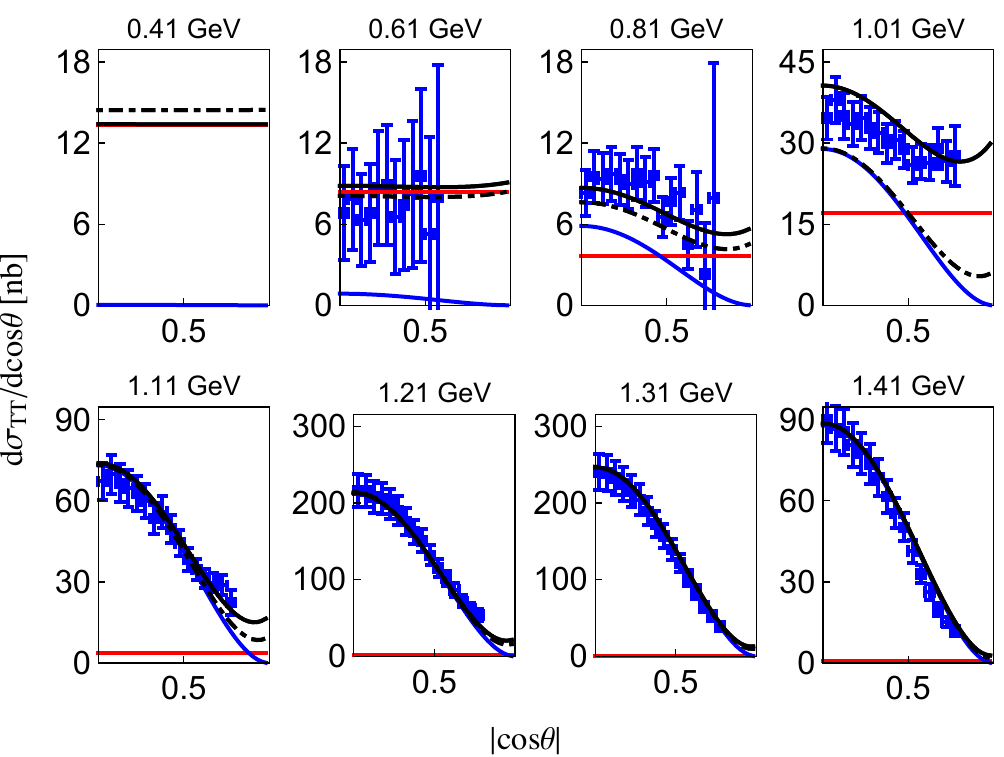}
\caption{Total and differential cross sections for $\gamma\gamma \to \pi^+\pi^-,\, \pi^0\pi^0$. 
The coupled-channel (single-channel) results are shown by the solid (dashed-dotted) black curves.  The separate contributions from the $s$-waves ($d$-waves) are shown by the red (blue) curves, whereas the Born result is shown by dashed gray curves. 
The data are taken from
\cite{Marsiske:1990hx,Uehara:2009cka,Boyer:1990vu,Behrend:1992hy,Mori:2007bu}.
\label{fig:Q2=0}}
\end{figure*}

The vector-meson exchange left-hand cuts are obtained by the effective Lagrangian which couples photon, vector (V) and pseudoscalar (P) meson fields,
\begin{equation}\label{LVPg}
{\cal  L}_{VP\gamma}=e\,C_V\,\epsilon^{\mu\nu\alpha\beta}\,F_{\mu\nu}\,\partial_\alpha P\,V_\beta\,,
\end{equation}
where $F_{\mu\nu}=\partial_\mu\,A_\nu-\partial_\nu\,A_\mu$. The PDG values \cite{Patrignani:2016xqp} for the partial decay widths
\begin{equation}
\Gamma _{V\to P\gamma}=\frac{e^2\,C_{V}^2\left(m_V^2-m_P^2\right)^3 }{24\,\pi\,m_V^3}
\end{equation}
allows to estimate the modulus of the radiative couplings by SU(3) relations as
\begin{equation}\label{gVP}
g_{V}\simeq C_{\rho^{\pm,0}}\simeq \frac{C_{\omega}}{3} \simeq \frac{1}{2} C_{K^{*0}} \simeq C_{K^{*\pm}}= 0.4(1) \,\text{GeV}^{-1} .
\end{equation}
In the following we will use $g_V$ as the only fit parameter, as discussed below, yielding $g_{V} =0.33$ GeV$^{-1}$, which is well in agreement with the SU(3) range of Eq.~(\ref{gVP}). 

The off-shellness of the photon can be taken into account by the vector transition form factor which is defined as
\begin{align}\label{def:FVi}
\langle V(k, \lambda)| j_{\mu}(0) | \pi(p)\rangle = 2e\,C_Vf_{V,\pi}(Q^2)\,\epsilon _{\mu \alpha \beta \gamma}k^{\alpha}\,p^{\beta }\epsilon^{\gamma *}(k, \lambda).
\end{align}
We obtain the following invariant amplitudes
\begin{align}\label{Fi:Vexch}
&F_1^{Vexch}=-\sum _V \frac{e^2\,C_{V}^2}{2} \left(\frac{4\,t+Q^2}{t-m_V^2}+(t\to u)\right)f_{V,i}(Q^2)\,,\quad \nonumber \\
&F_2^{Vexch}=\sum _V \frac{e^2\,C_{V}^2}{2} \left(\frac{1}{t-m_V^2}+\frac{1}{u-m_V^2}\right)f_{V,i}(Q^2)\,,\\
&F_3^{Vexch}=\sum _V \frac{e^2\,C_{V}^2}{t-u}\left(\frac{1}{u-m_V^2}-\frac{1}{t-m_V^2}\right)f_{V,i}(Q^2)\,.\nonumber 
\end{align}
As for the pion pole, one can show that the vector pole contribution corresponds to replacing $t$ and $u$ by $m_V^2$ in the numerators of Eq.~(\ref{Fi:Vexch}). We emphasize  that for the DRs written in the form (\ref{Rescattring:I=0}, \ref{Rescattring:general}) only $\text{Disc}\, h^{(J),Vexch}_{\lambda_1\lambda_2}(s)$ is required as input, which is unique for the vector-pole contribution. In addition, the discontinuity along the left-hand cut does not have any polynomial ambiguities \cite{GarciaMartin:2010cw} and is asymptotically bounded at high energy. We note, that for the single virtual case, the left-hand cut consist of two pieces: $(-\infty, s_L^{(-)}] $ and $[s_L^{(+)}, 0]$, with 
\begin{align}\label{s_L}
s_L^{(\pm)}&=\frac{1}{2}\left(2\,m_i^2-Q^2-m_V^2-\frac{m_i^2\,(m_i^2+Q^2)}{m_V^2}\right)\\
&\pm \frac{(m_V^2-m_i^2)\,\lambda^{1/2}(m_V^2,m_i^2,-Q^2)}{2\,m_V^2}\,,\nonumber
\end{align}
where $\lambda$ is the K\"all\'en triangle function. For the electromagnetic transition form factor of the $\omega$ we use the dispersive analysis from \cite{Danilkin:2014cra} (see also \cite{Schneider:2012ez}). For the other (sub-dominant) contributions from the vector mesons, we use the VMD model \cite{Sakurai1969}.

\begin{figure*}[t]
\centering
\includegraphics[width =0.37\textwidth]{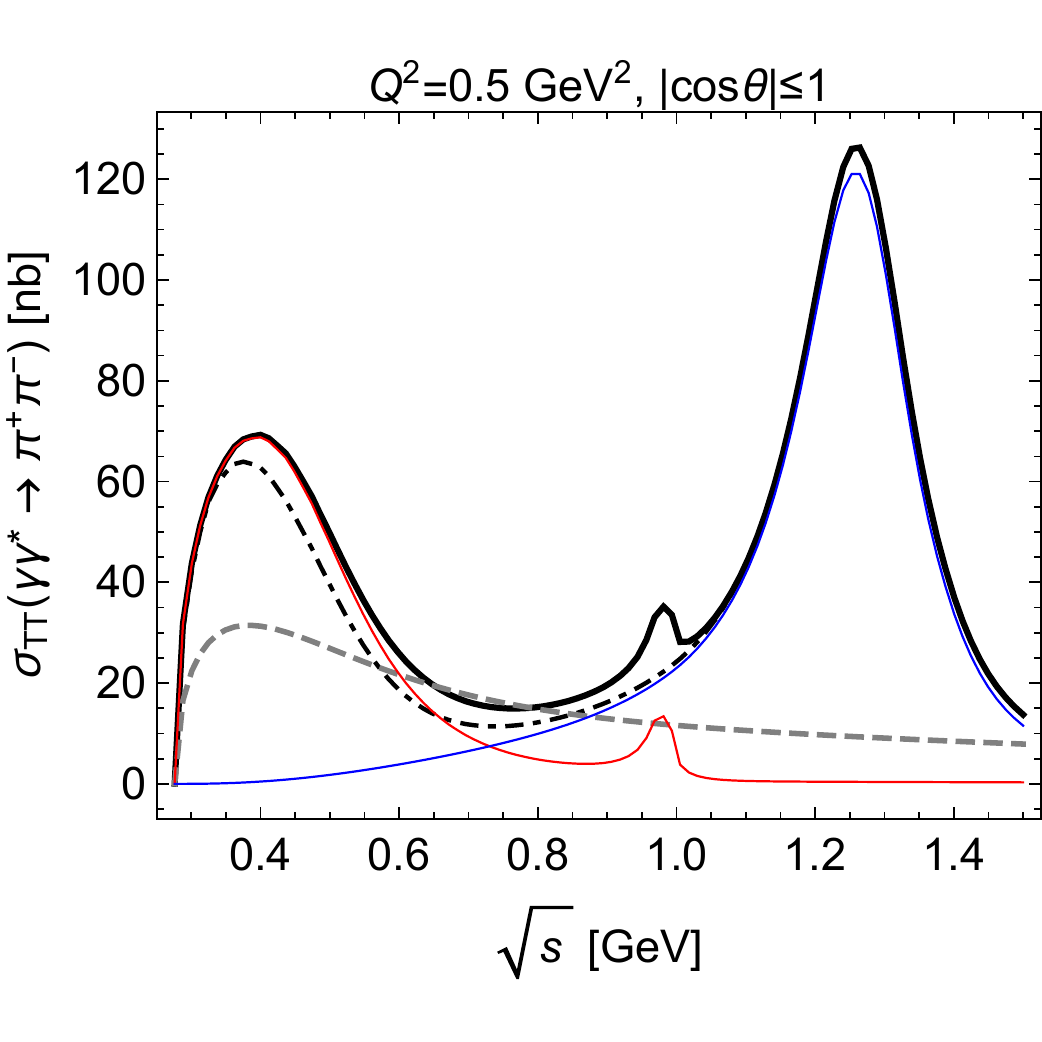}\quad \includegraphics[width =0.47\textwidth]{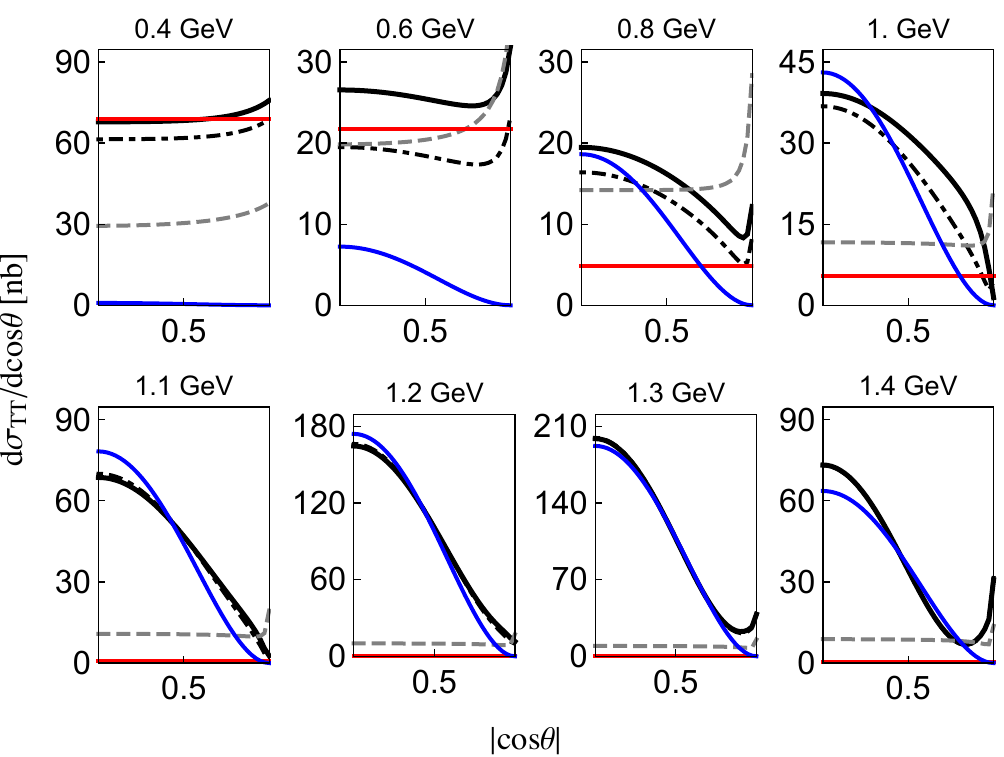}\\
\includegraphics[width =0.37\textwidth]{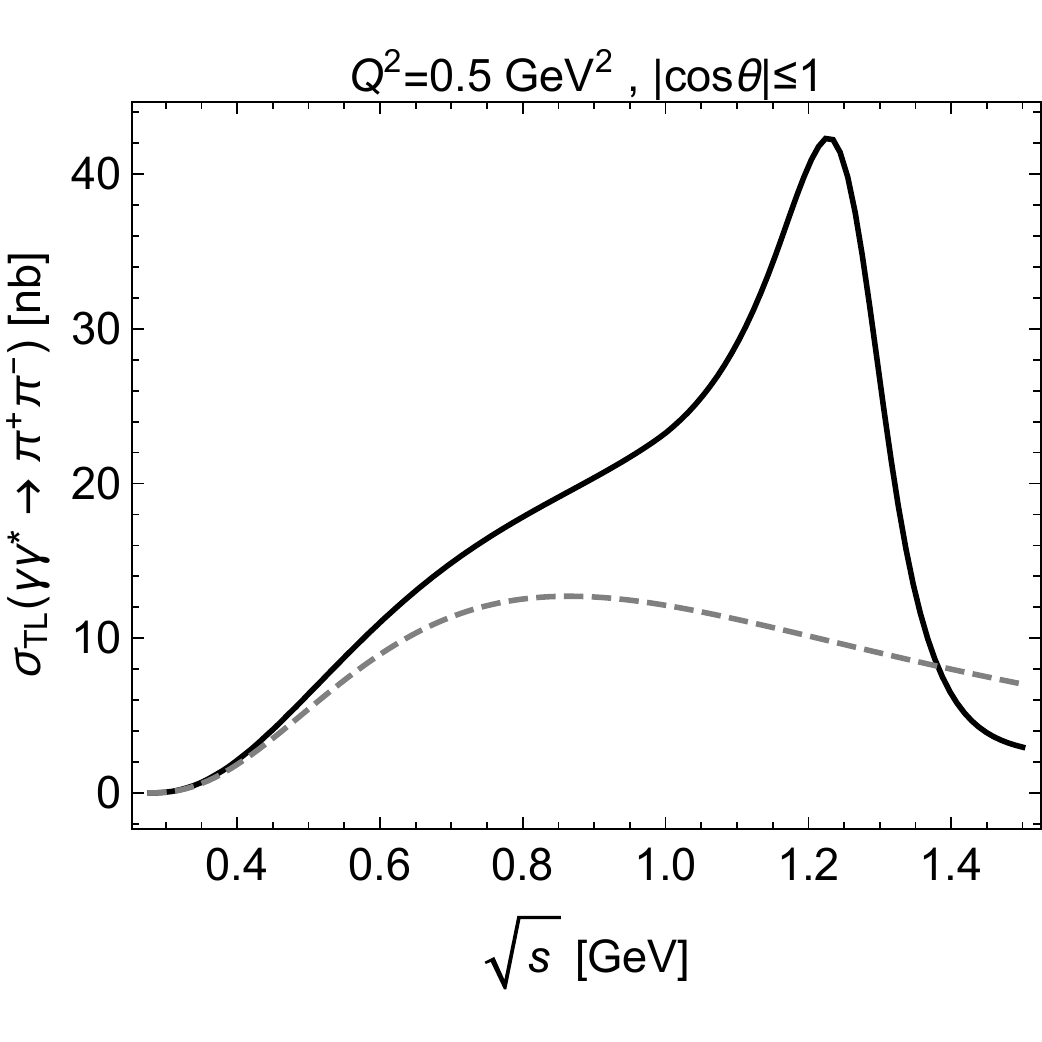}\quad \includegraphics[width =0.47\textwidth]{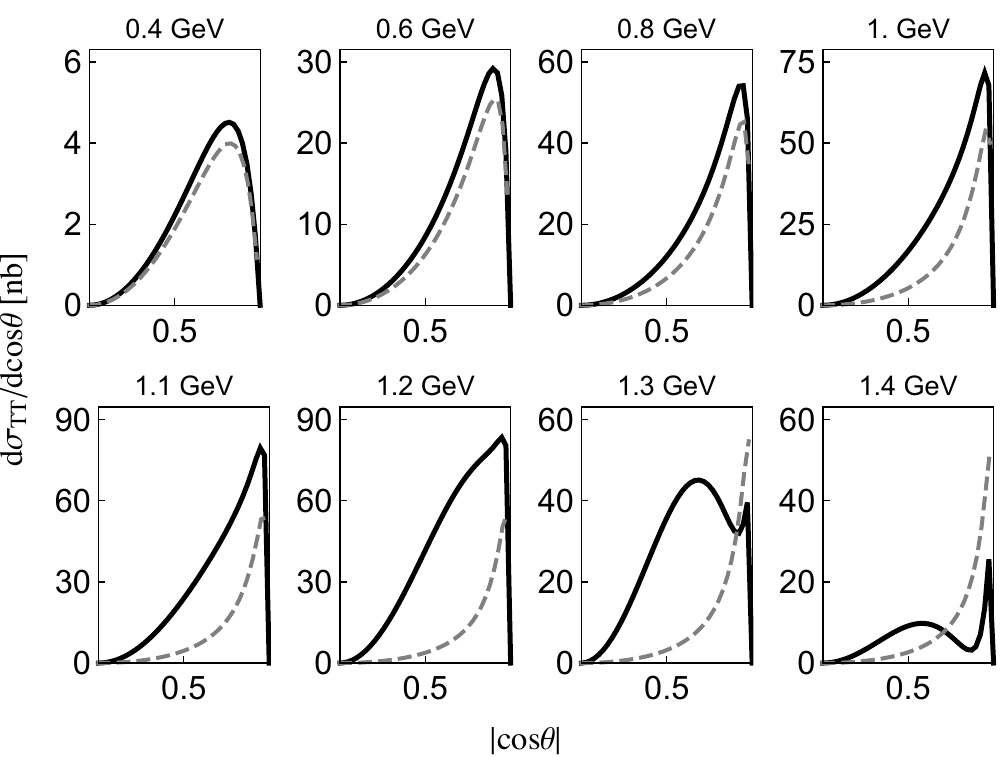}
\caption{Total and differential cross sections for $\gamma\gamma^* \to \pi^+\pi^-$ with $Q^2=0.5$ GeV$^2$ and full angular coverage $|\cos\theta| \leq 1$.
The coupled-channel (single-channel) results are shown by the solid (dashed-dotted) black curves.  The separate contributions from the $s$-waves ($d$-waves) are shown by the red (blue) curves, whereas the Born result is shown by dashed gray curves. 
\label{fig:Q2=0.5_charged}}
\end{figure*}

\begin{figure*}[t]
\centering
\includegraphics[width =0.37\textwidth]{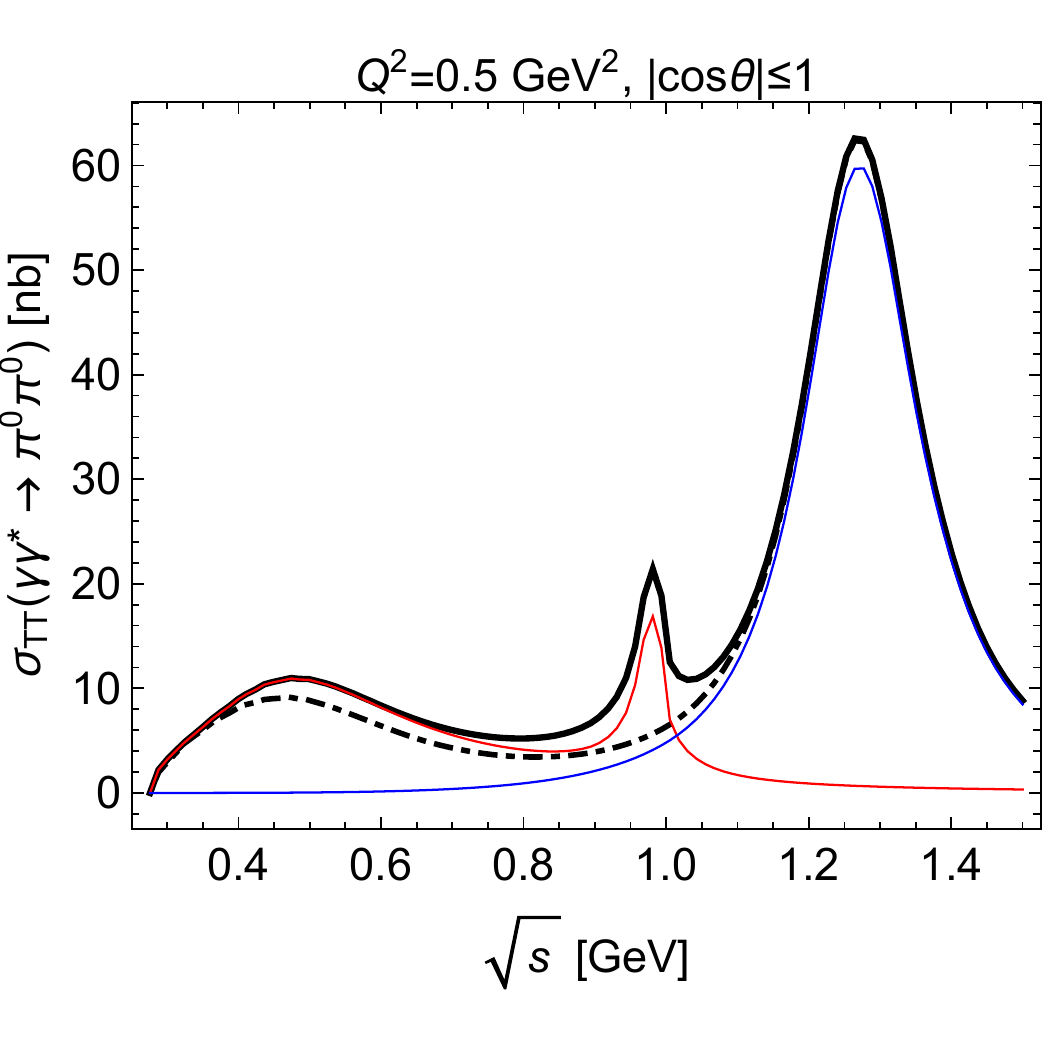}\quad \includegraphics[width =0.47\textwidth]{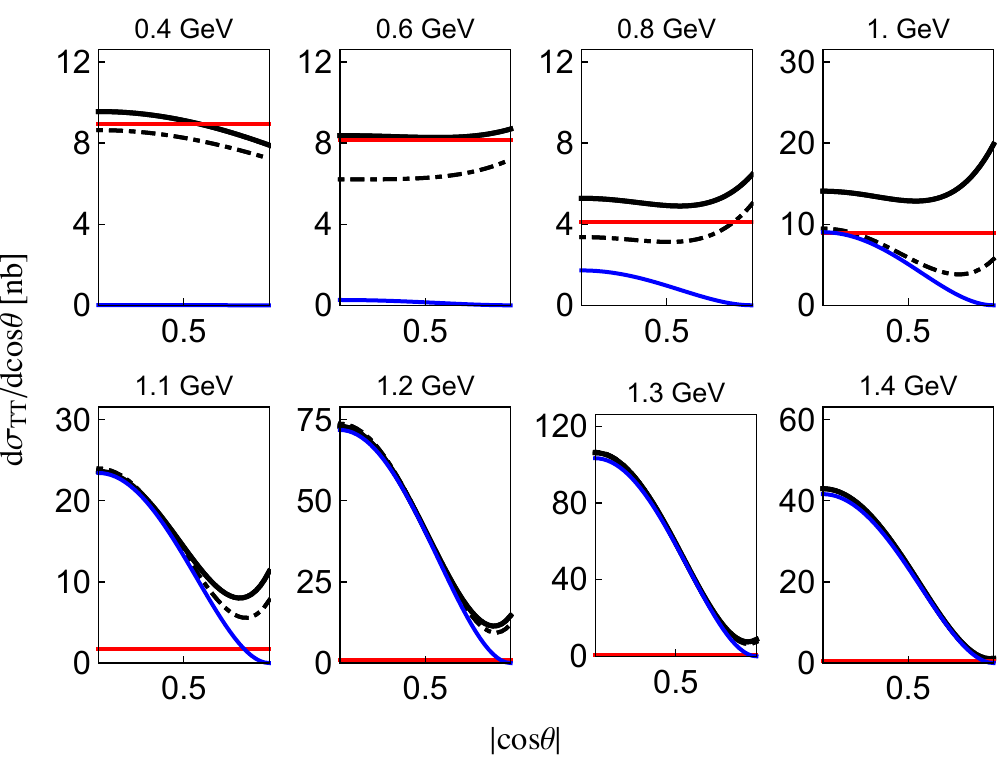}\\
\includegraphics[width =0.37\textwidth]{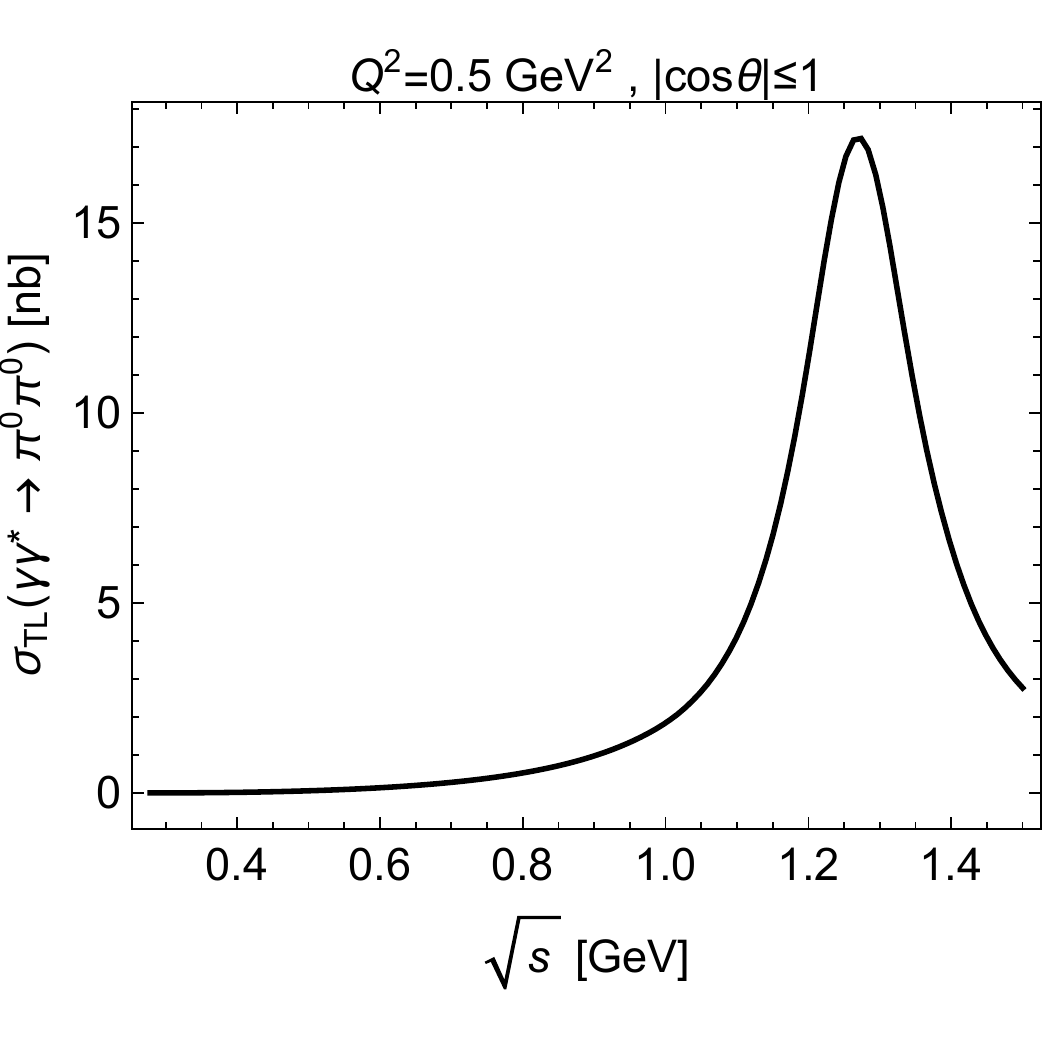}\quad \includegraphics[width =0.47\textwidth]{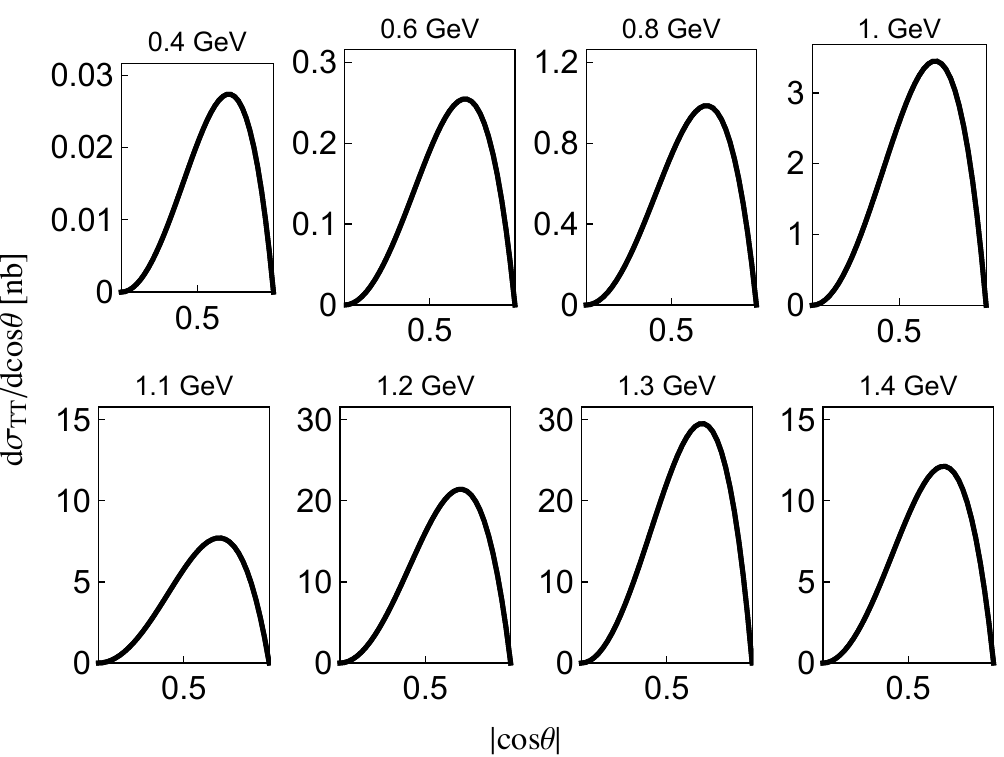}
\caption{Same as Fig.\ref{fig:Q2=0.5_charged} but for $\gamma\gamma^\ast \to \pi^0\pi^0$. 
\label{fig:Q2=0.5_neutral}}
\end{figure*}

\section{Numerical results}
\label{Section:Numerical results}
We start the discussion with the $s$-wave contribution. We find that the rescattering of the Born terms alone can be taken into account using the unsubtracted DR given in Eq.(\ref{Rescattring:I=0}). Results are shown in Fig.\ref{Fig:I=0} compared to the pure Born contribution. One can see that the coupled-channel analysis provides the narrow peak around 1 GeV which correspond to the $f_0(980)$ resonance. It can be compared to a single-channel case, where for $I=0$ we have used as input the phase shift from the single-channel inverse-amplitude method \cite{GomezNicola:2007qj} and Eq. (\ref{OmenesPhaseShift}), as it was done in \cite{Colangelo:2017fiz,*Colangelo:2017qdm,*Colangelo:2014dfa}. The details of the $f_0(980)$ resonance strongly depend on the input from the $\pi\pi \to K\bar{K}$ data which is not well known at present. Since the effect of $f_0(980)$ on the $\gamma\gamma^{\ast} \to \pi\pi$ channel is relatively small, we postpone a detailed analysis to a future work and for now take the result of the Omn\`es approach which is consistent with the current Roy analyses both for $\pi\pi \to \pi\pi$ \cite{Ananthanarayan:2000ht,GarciaMartin:2011cn} and for $\pi\pi \to K\bar{K}$ \cite{Buettiker:2003pp,Pelaez:2018qny} in the range till $1.2$ GeV. Using unsubtracted dispersive relations, we predict the following pion dipole polarizability as a check of the low energy limit: $(\alpha_1-\beta_1)_{\pi^\pm}= 6.1\,[5.5]\times 10^{-4}$ fm$^3$ where in brackets the single-channel result is indicated. This result is consistent with NNLO $\chi$PT $(\alpha_1-\beta_1)^{\chi PT}_{\pi^\pm}= 5.7(1.0)\times 10^{-4}$ fm$^3$ \cite{Gasser:2006qa} and with the recent COMPASS measurement:  $(\alpha_1-\beta_1)^{exp}_{\pi^\pm}=4.0(1.2)_{stat}(1.4)_{syst}\times 10^{-4}$ fm$^3$ \cite{Adolph:2014kgj}. The dipole polarizability for the neutral pion comes out as $(\alpha_1-\beta_1)_{\pi^0}= 9.5\, [8.9]\times 10^{-4}$ fm$^3$, which is far away from the NNLO $\chi$PT value of $(\alpha_1-\beta_1)^{\chi PT}_{\pi^0}= -1.9(0.2)\times 10^{-4}$ fm$^3$ \cite{Gasser:2005ud}. Similar results have also been observed in \cite{Colangelo:2017fiz,*Colangelo:2017qdm,*Colangelo:2014dfa}. The large value of the $\pi^0$ dipole polarizability is reflected in the absence of the Adler zero. Nevertheless, this mismatch to $\chi$PT is hardly visible on the $\gamma\gamma \to \pi^0\pi^0$ cross section, since its main contribution comes from the rescattering process $\gamma\gamma \to \pi^+\pi^-\to\pi^0\pi^0$. We note that the polarizabilities are saturated by $90\%$ from the dispersion integral over the low energies $<1.4\,\text{GeV}$. This is no longer the case of the generalized polarizabilities. For instance, for $Q^2=0.5\,\text{GeV}^2$, the generalized polarizabilities $(\alpha_1-\beta_1)_{\pi^\pm}= 0.86\times 10^{-4}$ fm$^3$ and $(\alpha_1-\beta_1)_{\pi^0}= 1.62\times 10^{-4}$ fm$^3$ are saturated by $70\%$ from the region $<1.4\,\text{GeV}$, indicating the importance of higher energies. The dipole polarizabilities for $\pi^0$ are expected to get large corrections once vector-meson left-hand cuts are added since they are much stronger for the neutral channel due to $\omega$-exchange. However, any Lagrangian-based field theory result has a bad high energy behavior and requires adding at least one subtraction in the DR to cure it. This reduces the predictive power of the DRs. With light vector mesons as additional left-hand cuts, the once-subtracted result is fixed to the COMPASS result for the $\pi^{\pm}$ and NLO $\chi$PT for the $\pi^0$ and $K$.
% (isospin zero). 
The comparison between unsubtracted and once-subtracted results is shown in Fig.\,\ref{Fig:I=0}. The comparison indicates a very similar description up to about 1.1 GeV.  Therefore, we decided to stay with the unsubtracted DR in the rest of this work, especially since the finite $Q^2$ prediction from $\chi$PT for the generalized polarizabilities are expected to be valid only in a very small $Q^2$ region.

While the contribution from the Born left-hand cut should be dominant at low energies (due to small pion mass in the t-channel), a description the $f_2(1270)$ region requires adding higher-mass intermediate states in the left-hand cuts \cite{GarciaMartin:2010cw}. We approximate them with vector-pole contributions. The radiative decay coupling $g_{V}$ in Eq.(\ref{gVP}) is fixed at the $f_2(1270)$ resonance position from the $\gamma\gamma\to \pi^0\pi^0$ cross section as discussed above. We emphasize that this is the only parameter that we adjust to the real photon data, within its expected range. The results for the differential and total cross section are shown in Fig.\,\ref{fig:Q2=0}. One can see that we achieve a reasonable description of the charged and neutral channels with the exception of the intermediate region in $\gamma\gamma \to \pi^+\pi^-$ and a slightly stronger $f_0(980)$. This in principle can be fixed by over-subtracting the DR and fitting this unknown subtraction constant to the data, similar to the analysis in \cite{GarciaMartin:2010cw}. However, our main goal in this work is to have a predictive power for the single virtual process. 

Our prediction for the spacelike single virtual case using the unsubtracted DR formalism is shown in Figs.\,\ref{fig:Q2=0.5_charged} and \ref{fig:Q2=0.5_neutral} for $\sigma_{TT}$ and $\sigma_{TL}$. The latter is fully determined by the helicity-1 contributions and increases with increasing $Q^2$ in the low $Q^2$ regime. For the $\sigma_{TT}$ we emphasize the importance of the unitarization, which increases the pure Born prediction at low energy by approximately a factor of two. Coupled-channel effects are important not only in the $f_0(980)$ region, increasing its importance of the future $a_\mu$ extraction. For $\sigma_{TL}$ we notice that the angular distribution is forward peaked due to the Born contribution. We defer a detailed discussion of error estimates to a forthcoming work \cite{Danilkin:2018}.

\section{Conclusions}

In this work, we have presented a dispersive analysis of the $\gamma\gamma^* \to \pi\pi$ reaction from the threshold up to 1.5 GeV in the $\pi\pi$ invariant mass. For the s-wave, we used a coupled-channel dispersive approach in order to adequately describe the scalar $f_0(980)$ resonance, which has a dynamical $\{\pi\pi, K\bar{K}\}$ origin. Since the s-wave provides a dominant contribution at low energy, we used only Born left-hand cuts. We have compared unsubtracted and subtracted DR formalisms and have shown that up to a c.m. energy of 1.1 GeV the subtraction does not change the results significantly, corroborating our choice of using an unsubtracted DR framework. For finite $Q^2$ we have demonstrated the importance of $K\bar{K}$ intermediate states for the first time.

Since $f_{2}(1270)$ tensor resonance decays predominantly to two pions, we have employed a single channel dispersive approach that requires $t$- and $u$-channel vector-meson exchange contributions to the left-hand cut. The only parameter in our approach is the $V P \gamma$ coupling which we fixed from the real photon data, and which is found to fall within the SU(3) spread between the couplings determined from experimental vector meson radiative decays. We achieved a reasonable description of the  $\gamma\gamma \to \pi^+\pi^-, \pi^0\pi^0$ total cross sections in comparison with the recent empirical data from the Belle Collaboration \cite{Uehara:2009cka,Mori:2007bu}. For the finite $Q^2$ we made a first dispersive prediction of the cross section including the $f_2(1270)$ region. Its measurement is part of an ongoing dedicated experimental program at BESIII.

The obtained results will serve as one of the relevant inputs to constrain the hadronic piece of the light-by-light scattering contribution to the muon's $a_\mu$~\cite{Colangelo:2017fiz,*Colangelo:2017qdm,*Colangelo:2014dfa,Pauk:2014rfa,Colangelo:2014pva}.

\section*{Acknowledgements}
%This work was supported by the Deutsche Forschungsgemeinschaft (DFG) in part through the Collaborative Research Center [The Low-Energy Frontier of the Standard Model (SFB 1044)], and in part through the Cluster of Excellence [Precision Physics, Fundamental Interactions and Structure of Matter (PRISMA)].

This work was funded by the Deutsche Forschungsgemeinschaft (DFG, German Research Foundation), in part through the Collaborative Research Center [The Low-Energy Frontier of the Standard Model,  Projektnummer 204404729 – SFB 1044] and in part through the Cluster of Excellence [Precision Physics, Fundamental Interactions and Structure of Matter (PRISMA)].

\bibliographystyle{apsrevM}
\bibliography{PhysLettB}

\end{document}